\documentclass[%
reprint, twocolumn, prb,
amsmath,amssymb,
aps,
]{revtex4}
\usepackage{graphicx}
\usepackage{dcolumn}
\usepackage{bm}
\usepackage{color}
\usepackage{multirow}
\usepackage{epsfig}
\usepackage{graphicx}
\usepackage{amsmath}
\usepackage{amssymb}
\usepackage{bm}
\usepackage{dcolumn}
\usepackage{braket}
\usepackage{bbold}
\usepackage{ulem}
\usepackage{titlesec}
\usepackage{lipsum}
\titlespacing*{\section}
{0pt}{0.5\baselineskip}{0.5\baselineskip}
\titlespacing*{\subsection}
{0pt}{0.5\baselineskip}{0.5\baselineskip}

\usepackage{etoolbox}
\BeforeBeginEnvironment{figure}{\vskip-0.0ex}
\AfterEndEnvironment{figure}{\vskip-0.0ex}

\textfloatsep{1pt plus 1pt minus 1pt}

\begin{document}

\preprint{APS/123-QED}

\title{Giant anomalous Hall and anomalous Nernst Conductivities in Antiperovskites and their Tunability via Magnetic Fields}

\author{Harish K. Singh}
\email{harish@tmm.tu-darmstadt.de}
\affiliation{Institute of Materials Science, Technical University Darmstadt, Otto-Berndt-Strasse 3, 64287 Darmstadt, Germany}

\author{Ilias Samathrakis}
\affiliation{Institute of Materials Science, Technical University Darmstadt, Otto-Berndt-Strasse 3, 64287 Darmstadt, Germany}
\author{Chen Shen}
\affiliation{Institute of Materials Science, Technical University Darmstadt, Otto-Berndt-Strasse 3, 64287 Darmstadt, Germany}
\author{Hongbin Zhang}
\email{hzhang@tmm.tu-darmstadt.de}
\affiliation{Institute of Materials Science, Technical University Darmstadt, Otto-Berndt-Strasse 3, 64287 Darmstadt, Germany}


\date{\today}

\begin{abstract}
The anomalous Hall conductivity (AHC) and anomalous Nernst conductivity (ANC) are two prominent transport phenomena in ferromagnetic materials of a topological nature. Based on first-principles calculations, we evaluated the AHC and ANC of 35 cubic ferromagnetic antiperovskites (APVs) and observed giant AHC and ANC as large as 1128 S/cm and 6.27 AK$^{-1}$m$^{-1}$ for Co$_3$LiN and Co$_3$PtN, respectively. 
Detailed analysis reveals that the origin of giant ANC can be attributed to the occurrence of Weyl nodes near the Fermi energy, as demonstrated for Co$_3$PtN.
Interestingly, both the magnitude and sign of AHC and ANC can be tuned by changing the magnetization ($\mathbf{M}$) directions, which could be applied to realize spin-caloritronics devices.     
\end{abstract}

\maketitle


\section{INTRODUCTION}
\begin{figure}
	\begin{center}
		\includegraphics[width=0.5\textwidth]{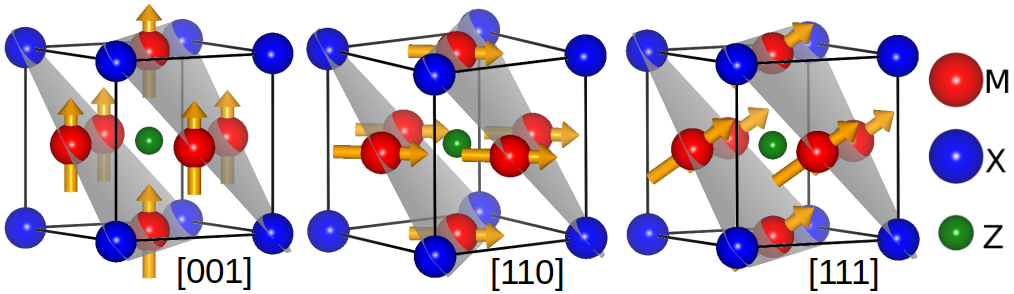}
		\caption{The crystal structure of antiperovskites with chemical formula M$_3$XZ in Pm$\bar{3}$m (221) space group, where M atoms are located at face-centers, whereas X atoms Z occupies the corner and body-centered position. The spin moments depict [001], [110], and [111] magnetization directions. }
		\label{fig:magS}
	\end{center}
\end{figure}

The anomalous Hall effect (AHE) is one of the emergent topological properties in ferromagnetic (FM) materials,~\cite{nagaosa2010anomalous} driven by the broken time-reversal symmetry and spin-orbital coupling (SOC).~\cite{suzuki2017cluster} 
Plenty of reported FM materials exhibit large AHC, such as CrPt$_3$ (2040 S/cm)~\cite{markou2020hard} and Co$_2$MnAl (1800 S/cm)~\cite{kubler2012berry}. 
Moreover, the quantum AHE is interesting for spintronic applications such as magnetic sensors, memory devices, and data processing.~\cite{he2018topological, nadeem2020quantum, vzutic2004spintronics, hirohata2020review}  
Recently, it was demonstrated that the AHC hinges on the magnetization direction.~\cite{roman2009orientation,zhang2011anisotropic} 
For example, the AHC of hcp Co decreases from 481 to 116 S/cm as magnetization direction changed from c-axis (easy-axis) to the ab plane, respectively.~\cite{roman2009orientation} Whereas for CoPt, both AHC sign and magnitude altered, the AHC change from -119 to 107 S/cm, switching magnetization [001] to [110] direction, indicating a change in the AHC sign.~\cite{zhang2011anisotropic} From a symmetry point of view, the magnetic space group of the material is closely related to the magnetic spin directions, which leads to change band topology and hence likely change in magnitude and sign of AHC appear. Besides FM, finite AHC has been recently observed in noncollinear antiferromagnets,~\cite{chen2014anomalous} emerging from their exotic antiferromagnetic (AFM) spin order ($\Gamma_{4g}$/$\Gamma_{5g}$) in cubic antiperovskites Cr$_3$XN (X = Ir and Pt) and Mn$_3$XN (X = Ag, Au, Co, Ga, Hg, In, Ir, Ni, Pd, Pt, Rh, and Zn)~\cite{huyen2019topology,singh2020multifunctional,boldrin2019anomalous,Gurung,zhou2020giant,zhou2019spin,Samathrakis-Mn3GaN}, cubic Mn$_3$X (Z = Ir, Pt, and Rh), and hexagonal Mn$_3$X (X = Ga, Ge, and Sn).~\cite{guo2017large,chen2014anomalous, kubler2014non,nayak2016large,kiyohara2016giant,zhang2017strong}      
\\
\\
Recently, the energized field of spin-caloritronics focuses on the correlation between spintronics and thermoelectric in magnetic materials, attracted considerable attention due to their potential application in thermoelectric devices.~\cite{slachter2010thermally,bauer2012spin,boona2014spin} More interestingly, the concomitant anomalous Nernst effect (ANE) indicates that an electric current can be generated perpendicular to both the applied heat gradient and magnetization direction.~\cite{mizuguchi2019energy,behnia2016nernst} 
In this context, Weyl semimetals are promising to induce a significant ANE, owing to the occurrence of Weyl nodes near the Fermi-energy.~\cite{liu2018giant,sakai2018giant,xu2021unconventional}
For instance, a large anomalous Nernst conductivity 
(ANC) of 10.0 AK$^{-1}$m$^{-1}$ was reported for the Weyl semimetal Co$_3$Sn$_2$S$_2$, primarily originating from the large Berry curvature around the Weyl points (WPs) near the Fermi-energy.~\cite{liu2018giant,yang2020giant} 
Note that the number of the Weyl nodes can be tailored by magnetization direction hence has a sizeable impact on the electronic state's topological character,~\cite{zhang2018magnetization,zheng2019tunable} resulting in tuning the material's topological properties. Furthermore, the kagome lattice is one of the determinants to induce enhanced ANC.~\cite{asaba2021colossal,liu2018giant}
From this perspective, antiperovskites are fascinating, host a kagome lattice in the (111) plane, and also, the Weyl nodes can be tuned with applied biaxial strain as observed in noncollinear Mn$_3$PdN.~\cite{singh2020multifunctional}  
\\
\\
In this work, we systematically evaluated the AHC and ANC of 35 cubic ferromagnetic APV compounds M$_3$XZ (Table~\ref{tab:table1} and Figure~\ref{fig:magS}). All 35 APVs considered in this study fulfill thermodynamical, mechanical, and dynamical stability criteria, comprising 24 experimentally reported compounds and 11 APVs predicted from our high-throughput screening.~\cite{Singh2018} The APVs in FM state exhibit a large AHC and ANC, such as Co$_3$LiN attains an absolute AHC as large as 1128 S/cm, and Co$_3$PtN displays a giant ANC of 6.27 AK$^{-1}$m$^{-1}$, which is nearly the same as in cubic Fe$_3$Pt ferromagnet (6.2 AK$^{-1}$m$^{-1}$).~\cite{sakai2020iron} We investigated the Weyl points (WPs) to identify the origin of large AHC and ANC, revealing the WPs contribution to the total AHC and ANC. Next, we carried out a comparison of AHC and ANC between FM and noncollinear APVs. Thus, we distinguished that the FM APVs have larger AHC and ANC than the noncollinear APVs. Further, we realized that the AHC and ANC strongly depend on the magnetization directions and can be tuned significantly by switching the magnetization direction. In this context, we presented a thermopile structure to explain the comprehensive interplay between electric current ($\mathbf{I}$), thermal gradient ($\nabla$T), and magnetization ($\mathbf{M}$) directions.
\section{COMPUTATIONAL DETAILS}
Our density functional theory (DFT) calculations are conducted using the projector augmented wave (PAW) method as implemented in the VASP code.~\cite{kresse1996} A generalized gradient approximation (GGA) is used within the parametrized form of Perdew-Burke-Ernzerhof
(PBE) to treat exchange-correlation functional.~\cite{perdew1996generalized} The energy cutoff of 600 eV is used for the plane-wave basis set, and the spin-orbit coupling (SOC) is taken into account for each APV system. A uniform k-mesh of 17$\times$17$\times$17 is employed for the Brillouin zone integrations within the Monkhorst-pack scheme. The partial occupancies of the electronic states are smeared by using the Methfessel-Paxton smearing width of 0.06 eV. The total energy convergence criterion is set to 10$^{-07}$. The optimized lattice constants are considered from our previous high throughput study.~\cite{Singh2018}
\\
We used the Wannier90 code to obtain the maximally localized Wannier functions (MLWFs) and get the tight-binding model Hamiltonian.~\cite{mostofi2008wannier90} In total, 80 MLWFs are constructed for every APV system by projecting the s, p, and d orbitals of M and X atoms and the  s and p orbitals for the N or C atom. The AHC is computed using the WannierTools code.~\cite{Wu2017} A uniform k-mesh of 401$\times$401$\times$401 is used for the Berry curvature integration.~\cite{xiao2010berry} The AHC is determined in conformity with the following equation:

\begin{equation}
\label{eq:AHC}
\sigma_{\alpha\beta} = - \frac{e^2}{\hbar} \int \frac{d\mathbf{k}}{\left( 2\pi \right)^3} \sum_{n} f \big[ \epsilon \left( \mathbf{k} \right) -\mu \big] \Omega_{n,\alpha\beta} \left( \mathbf{k} \right) 
\end{equation} 
\begin{equation}
\label{eq:Berry}
\Omega_{n,\alpha\beta}\left( \mathbf{k} \right) = -2\text{Im} \sum_{m \neq n} \frac{{\braket{\psi_{\mathbf{k}n}|v_{\alpha}|\psi_{\mathbf{k}m}}}{\braket{\psi_{\mathbf{k}m}|v_{\beta}|\psi_{\mathbf{k}n}}}}{\big[ \epsilon_m\left( \mathbf{k} \right) - \epsilon_n\left( \mathbf{k} \right) \big]^2} 
\end{equation}
where e is elementary charge, $\mu$ is the chemical potential, $\psi_{n/m}$ denotes the Bloch wave function with energy eigenvalue $\epsilon_{n/m}$,
$v_{\alpha/\beta}$ is the velocity operator along Cartesian $\alpha/\beta$ direction, and $f[\epsilon \left( \mathbf{k} \right) -\mu]$ is Fermi-Dirac distribution function. The ANC is computed using the in-house developed python code. The validation of code is done by reproducing the ANC of Mn$_3$NiN a noncollinear APV~\cite{zhou2020giant} (Figure S1). 
The ANC is determined using the following equation: 
\begin{equation}
\label{eq:ANC}
a_{xy} = -\frac{1}{e} \int d\epsilon \frac{\partial f}{\partial \mu} \sigma_{\alpha \beta} \left(\epsilon\right) \frac{\epsilon - \mu}{T},
\end{equation}
where T is the temperature, $\mu$ the Fermi level, and $\epsilon$ is the point of the energy grid.
\\
\begin{table*}
	\caption{\label{tab:table1}The Compilation of MAE, AHC, and ANC of the APV compounds. The MAE is calculated using equation ~\ref{eq:MAE}. The absolute values of AHC and ANC are obtained using these using equations, $\sigma$$_{[001]}$=$\sigma$$_{z}$, $\sigma$$_{[110]}$=$\dfrac{1}{\sqrt{2}}$ ($\sigma$$_{x}$+$\sigma$$_{y}$) and $\sigma$$_{[111]}$=$\dfrac{1}{\sqrt{3}}$($\sigma$$_{x}$+$\sigma$$_{y}$+$\sigma$$_{z}$), AHC and ANC are presented at the Fermi-energy, and the ANC data is provided for 300 K. The energy-dependent AHC and ANC plots are provided in supporting information figure S3 and S4). The ANC plots as a function are summarized in the figure S5.}
	\begin{ruledtabular}
		\begin{tabular}{c|c|ccc|ccc|ccc}
			&M$_3$XZ&\multicolumn{2}{c}{MAE ($\mu$eV/atom)}&&\multicolumn{3}{c}{AHC (S/cm)}&\multicolumn{3}{c}{ANC (A/mK)}\\
			S.No.&Compounds&[001-111]&[001-110]&[110-111]&[001]&[110]&[111]&[001]&[110]&[111]\\ \hline\centering
			(1) &Co$_3$AuN&-4.4&	-4.2 &-0.2&    480.2 &    536.3 &	549.0 &  0.97 & -0.65 & -0.37 \\
			(2) &Co$_3$CdN&-0.2 &0.8 &-1.0 &785.9 &753.2 &   604.9 & -3.27 & -3.39& -3.93 \\
			(3) &Co$_3$CuN&   6.6 &	     -0.8 &7.4&    300.9 &    572.3 &   591.3 &  2.84 &  3.32 &  2.35 \\
			(4) &Co$_3$GaN&   -0.8 &	-1.4 &0.6 &    557.2 &    530.6 &   536.7 & -0.58 & -1.34 &  0.05 \\
			(5) &Co$_3$GeC&  -8.0 &	-11.4 &3.4&     72.9 &    129.3 &	152.9 & -0.87 & -1.36 & -1.20 \\
			(6) &Co$_3$GeN&  -3.4 &	-2.8 &-0.6&    231.7 &    332.1 &	348.8 & -1.24 & -2.31 & -3.35\\ 
			(7) &Co$_3$InN&  -3.8 &	-5.0 &1.2&    482.6 &    476.0 &	470.5 & -0.57 & -0.01 &  0.66 \\
			(8) &Co$_3$LiN&   12.0 &  -42.4 &54.4&878.8&    166.0 &1128.7& -1.70 &  0.55 & -3.02 \\
			(9) &Co$_3$MgC&   -1.2 &	     -1.0 &-0.2&    283.4 &    289.5 &  -291.1 & -0.58 & -0.20 &  0.05 \\
			(10)&Co$_3$PdN&   -1.0 &	-2.6 &1.6&    158.6 &    191.1 &	242.1 &  0.94 &  0.48 &  0.26 \\
			(11)&Co$_3$PtN& -41.6 &  -33.0 & -8.6 &    710.4 &   559.9 &   760.5 &  6.27 &  4.41 &  3.79 \\
			(12)&Co$_3$RhN&   6.0 &  4.8 &1.2&   -116.1 &    244.3 &   258.9 & -0.46 &  0.51 &  0.54 \\
			(13)&Co$_3$SnC&   1.2  &   25.2 &-24.0&    -51.77 &    166.7 &   223.2 & -0.79 & -1.13 & -1.47 \\
			(14)&Co$_3$SnN&   3.0  &   -19.0 &22.0&     -0.0 &     32.3 &	48.1 & -2.58 & -2.75 & -2.37 \\
			(15)&Co$_3$ZnC& -0.8&-0.6&-0.2&    209.4 &   -281.5 &   266.7 & -0.38 &  0.02 &  0.15 \\
			(16)&Co$_3$ZnN&   0.8  &	4.6 &-3.8&    930.4&   1042.5 &  1069.6 & -3.23 &-4.41 & -4.79 \\
			(17)&Fe$_3$AgN&   -3.6 &-5.0&1.4&    189.4 &    365.9 &   318.1 &  1.17 &  2.20 &  2.41 \\
			(18)&Fe$_3$AlC&   1.0  &	 0.6 &0.4&   -135.8 &      0.2 &	33.7 &-0.025       &0.019&-0.105\\
			(19)&Fe$_3$AuN&  -4.2  &	-1.0 &-3.2&    514.7 &    603.0 &   626.9 &  1.20 &  0.42 &  1.08 \\
			(20)&Fe$_3$CuN&   3.4  &	 6.8 &-3.4&    630.1 &    845.8 &   891.6 & -0.08 &  1.31 &  1.49 \\
			(21)&Fe$_3$GaN& -1.8  &	 -1.2 &0.6&    262.2 &    242.7 &   215.6 &  2.55 &  2.68 &  2.65 \\
			(22)&Fe$_3$InN&  1.0  &	 0.8 &0.2&    491.1 &    408.3 &   374.6 &  0.33 &  0.47 &  0.92 \\
			(23)&Fe$_3$IrN& -59.8&  -59.2 &-1.6 &    841.9 &   721.2 &   932.1& -0.60 & -2.21 & -2.14 \\ 
			(24)&Fe$_3$NiN&  -1.2  &	 0.2 &-1.4&    286.1 &    402.9 &	29.5 &  0.02 & -0.34 &  0.12 \\
			(25)&Fe$_3$PdN& 52.0&	 0.6 &51.4 &    361.9 &    518.5 &   560.1 & -0.21 & -0.51 &  0.02 \\
			(26)&Fe$_3$PtN& -3.8  &	-0.8 &-3.0&   -401.6 &   -405.2 &  -523.7 &  1.34 &  0.68 &  0.98 \\
			(27)&Fe$_3$RhN& -3.6  &	-0.4 &-3.2&    170.6 &    124.8 &   134.4 & -1.39 & -2.00 & -2.76 \\
			(28)&Fe$_3$SnN& -2.4  &	-1.6 &-0.8&    123.9 &    162.8 &   185.1 & -1.61 & -0.90 & -1.02 \\
			(29)&Fe$_3$ZnC&   0.4  &	-1.2 &1.6&     63.5 &    119.9 &   182.3 & -0.77 & -1.39 & -0.77 \\
			(30)&Fe$_3$ZnN&-3.6&7.2&3.8&    -94.2 &   -82.4 &-14.4& -1.31 & -0.39 &0.26 \\
			(31)&Mn$_3$AlC&  -0.4  &	 1.2 &-1.6&    199.4 &    190.8 &   203.5 & -0.51 & -0.76 & -0.83 \\
			(32)&Mn$_3$AlN& -2.8  &	-0.4 &-2.4&     51.4 &     73.2 &	69.7 & -1.12 & -1.05 & -1.13 \\
			(33)&Mn$_3$InC& -8.6  &	-1.0 &-7.6&    434.9 &    290.4 &   235.8 &  1.96 &  1.79 &  1.56 \\
			(34)&Ni$_3$CuC&  5.2  &	4.0 &1.2&    254.4 &   -250.1 &   268.8 &  1.05 & -1.73 &  2.01 \\
			(35)&Ni$_3$LiC&  6.2  &	5.2 &1.0&    266.2 &    260.7 &   249.2 &  4.06 &  5.40&  5.80 \\
		\end{tabular}
	\end{ruledtabular}
\end{table*}
\begin{figure}
	\begin{center}
		\includegraphics[width=0.5\textwidth]{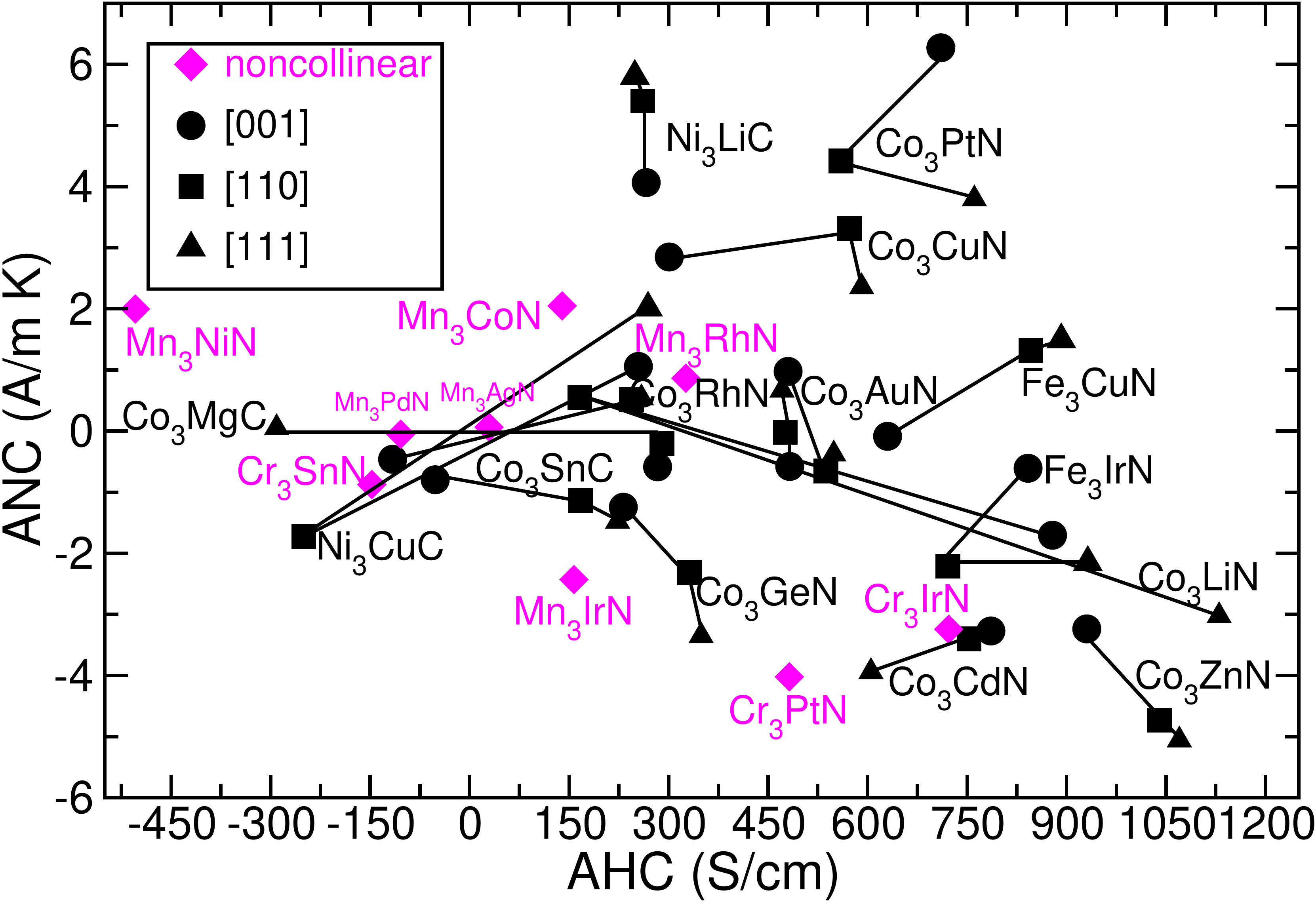}
		\caption{The comparison of AHC and ANC for the FM and noncollinear APV configurations. The solid line is adjoining [001], [110], and [111] magnetization directions for the individual FM APVs. The absolute values of AHC and ANC are illustrated at the Fermi-energy, and the ANC data is provided for 300 K.}
		\label{fig:AHC-ANC}
	\end{center}
\end{figure} 
\section{RESULTS and DISCUSSION}
\subsection{Magnetocrystalline Anisotropy Energy}
MAE is expressed as the difference of total energies for different magnetization directions, which yields
\begin{equation}
\label{eq:MAE}
\text{MAE}= \text{E}_{\mathbf{e}_1}-\text{E}_{\mathbf{e}_2}
\end{equation}
where $\mathbf{e}_1$ and $\mathbf{e}_2$ are chosen to be along the [001], [110], and [111] directions for the 
cubic APV compounds considered in this work. The resulting MAEs are shown in Table~\ref{tab:table1}. 
For cubic materials, it is well-known that MAE is mostly about a few $\mu$eV/atom, {\it e.g.}, the MAE of bcc Fe is -0.4 $\mu$eV/atom favoring the [001] direction as an easy axis.~\cite{halilov1998magnetocrystalline} 
It is found that the MAEs of 28 out of 35 cubic APVs are lower than 10 $\mu$eV/atom (Table~\ref{tab:table1}). Nevertheless, considerable MAE is obtained for Co$_3$LiN, Co$_3$PtN, Co$_3$RhN, Fe$_3$IrN, and Fe$_3$PdN (Table~\ref{tab:table1}). For instance, Fe$_3$IrN exhibits a significant MAE as large as -59.8 $\mu$eV with an easy axis in the [001] direction.
\\
\\
The magnitude of MAEs can be understood based on the strength of the atomic SOC, as marked by the spin-orbit coupling energy (E$_\text{SOC}$).~\cite{antropov2014constituents}
Taking Fe$_3$IrN as an example, ($\Delta$E$_\text{SOC}$) between the [001] and [111] magnetization directions read 
\begin{equation}
\label{eq:SOC}
\Delta\text{E$_\text{SOC}$}= \text{E}_\text{SOC} [001]-\text{E}_\text{SOC}[111],
\end{equation}
with the resulting $\Delta$E$_\text{SOC}$ as large as -295.1 $\mu$eV/atom. 
It indicates that the magnetization along the [001] direction is favored. 
Further analysis on the atom-resolved contributions reveals that 
the Ir atoms have dominant contributions (-1442 $\mu$eV/atom) in comparison 
to that ( -12.9 $\mu$eV/atom) of Fe atoms.
This can be attributed to the enhanced strength of atomic SOC, {\it i.e.}, $\xi_{5d}^\text{Ir}\sim$457 meV
in comparison to $\xi_{3d}^\text{Fe}\sim$54 meV.
The E$_\text{SOC}$ analysis is consistent for all the APV systems and equally valid to characterize their preferred magnetization direction.
\hspace*{-0.5cm}
\begin{figure*}
	\begin{center}
		\includegraphics[width=0.8\textwidth]{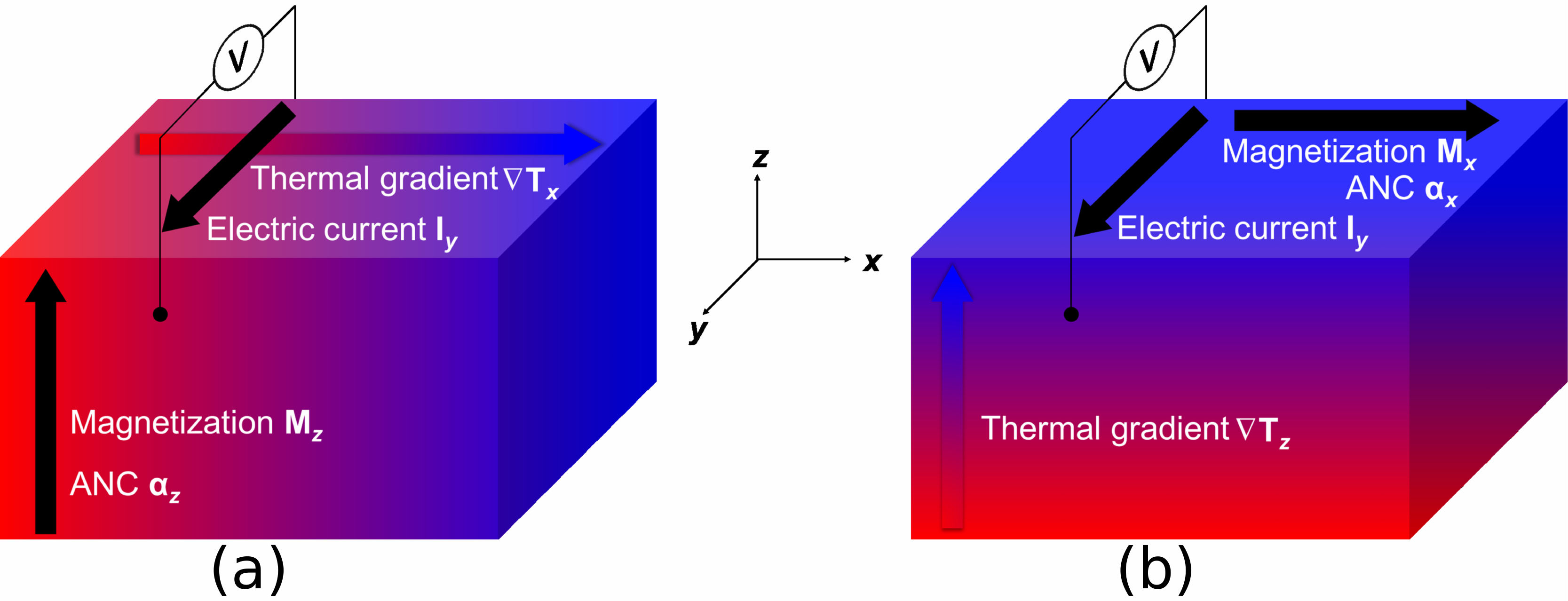}
		\caption{Schematics illustration of thermopile structure for the (a) applied in-plane thermal gradient ($\nabla$T$_x$) and Magnetization (M$_z$), generating an electric current in the y-direction and (b) exerted out-of-plane $\nabla$T$_z$ and M$_x$ also lead to an electric current in the y-direction.}
		\label{fig:Thermopile}
	\end{center}
\end{figure*}
\subsection{Symmetry analysis}
Symmetry plays an essential role in determining the linear response properties, including AHC and ANC. The shape of the AHC/ANC tensor elements can be obtained from the magnetic space group, where 
the resulting symmetry operations can lead to a specific distribution of the Berry curvature in the \textbf{k}-space and hence vanishing/finite values for AHC/ANC. For instance, the magnetic space group for the magnetization aligned along the [001], [110], and [111] directions are P4/mm$'$m$'$ (123.345), Cmm$'$m$'$ (65.486), and R-3m$'$ (166.101), respectively. For the [001] magnetization  direction, the Berry curvatures transform as follows: 
\begin{align}
&\Omega_x\left(-k_x,-k_y,-k_z\right)=-\Omega_x\left(k_x,k_y,k_z\right) \notag \\
&\Omega_y\left(-k_x,-k_y,-k_z\right)=-\Omega_y\left(k_x,k_y,k_z\right) \notag \\
&\Omega_z\left(-k_x,-k_y,-k_z\right)=\Omega_z\left(k_x,k_y,k_z\right) \notag.
\end{align}
Therefore, the resulting $\sigma_x$ and $\sigma_y$ components vanish by integrating the Berry curvatures following Eq.~\ref{eq:AHC}, which can be attributed to the two-fold rotation and mirror plane M$_{001}$ symmetry. Whereas the $\sigma$$_z$ component is invariant under both symmetry operations, giving rise to a finite AHC. 
This is confirmed by our explicit evaluation of AHC for Fe$_3$CuN as shown in Figure S2, with vanishing
$\sigma$$_x$ and $\sigma$$_y$ but a $\sigma$$_z$ as large as 630.1 S/cm.  
Similarly, for the [110] magnetization  direction, due to the M$'$$_{100}$ symmetry,$\Omega$$_z$(-k$_x$,-k$_y$,k$_z$)=-$\Omega$$_z$(k$_x$,k$_y$,k$_z$) resulting in vanishing $\sigma_z$,
while $\sigma$$_x = \sigma$$_y$ with nonzero values. Finally, for the [111] magnetization direction, the AHC components do not vanish under any symmetry operation such as Berry curvature is invariant for $\sigma$$_x$ components $\Omega$$_x$(k$_x$,k$_y$,k$_z$)=$\Omega$$_x$(k$_x$,k$_y$,k$_z$)  and in the same way for the corresponding $\sigma$$_y$ and $\sigma$$_z$ components. As a result, $\sigma_x = \sigma_y=\sigma_z$ within general finite values. Therefore, considering AHC as a pseudo-vector $\mathbf{\sigma}=\sigma_x\mathbf{e}_i+\sigma_y\mathbf{e}_j+\sigma_z\mathbf{e}_k$ (with $\mathbf{e}_{i/j/k}$ being the unit vectors along the Cartesian directions), it can be concluded that the AHC is always aligned with the magnetization direction for high-symmetric [001], [110], and [111] directions, as observed for tetragonal $3d$Pt alloys.~\cite{zhang2011anisotropic} Hereafter, we adopt $\sigma$$_{[001]}$=$\sigma$$_{z}$, $\sigma$$_{[110]}$=$\dfrac{1}{\sqrt{2}}$ ($\sigma$$_{x}$+$\sigma$$_{y}$) and $\sigma$$_{[111]}$=$\dfrac{1}{\sqrt{3}}$($\sigma$$_{x}$+$\sigma$$_{y}$+$\sigma$$_{z}$), as summarized in Table~\ref{tab:table1}, where, the various component of AHC are defined as $\sigma_{x}=\sigma_{yz}$, $\sigma_{y}=\sigma_{zx}$ and $\sigma_{z}=\sigma_{xy}$. Such symmetry analysis applies to ANC as well. For instance, a possible thermopile structure mechanism summarizing the interplay between thermal gradient ($\nabla$T), magnetization ($\mathbf{M}$), and electric current ($\mathbf{I}$) is depicted in Fig.~\ref{fig:Thermopile}. For an applied in-plane $\nabla$T$_x$ and out of plane magnetization $\mathbf{M}$$_z$, an electric current will generate along the y-direction (Fig.~\ref{fig:Thermopile}(a)), and for out-of-plane ($\nabla$T$_z$) and in-plane ($\mathbf{M}$$_x$) will also result in an electric current in the y-direction (Fig.~\ref{fig:Thermopile}(b)).
\begin{figure}
	\begin{center}
		\includegraphics[width=0.5\textwidth]{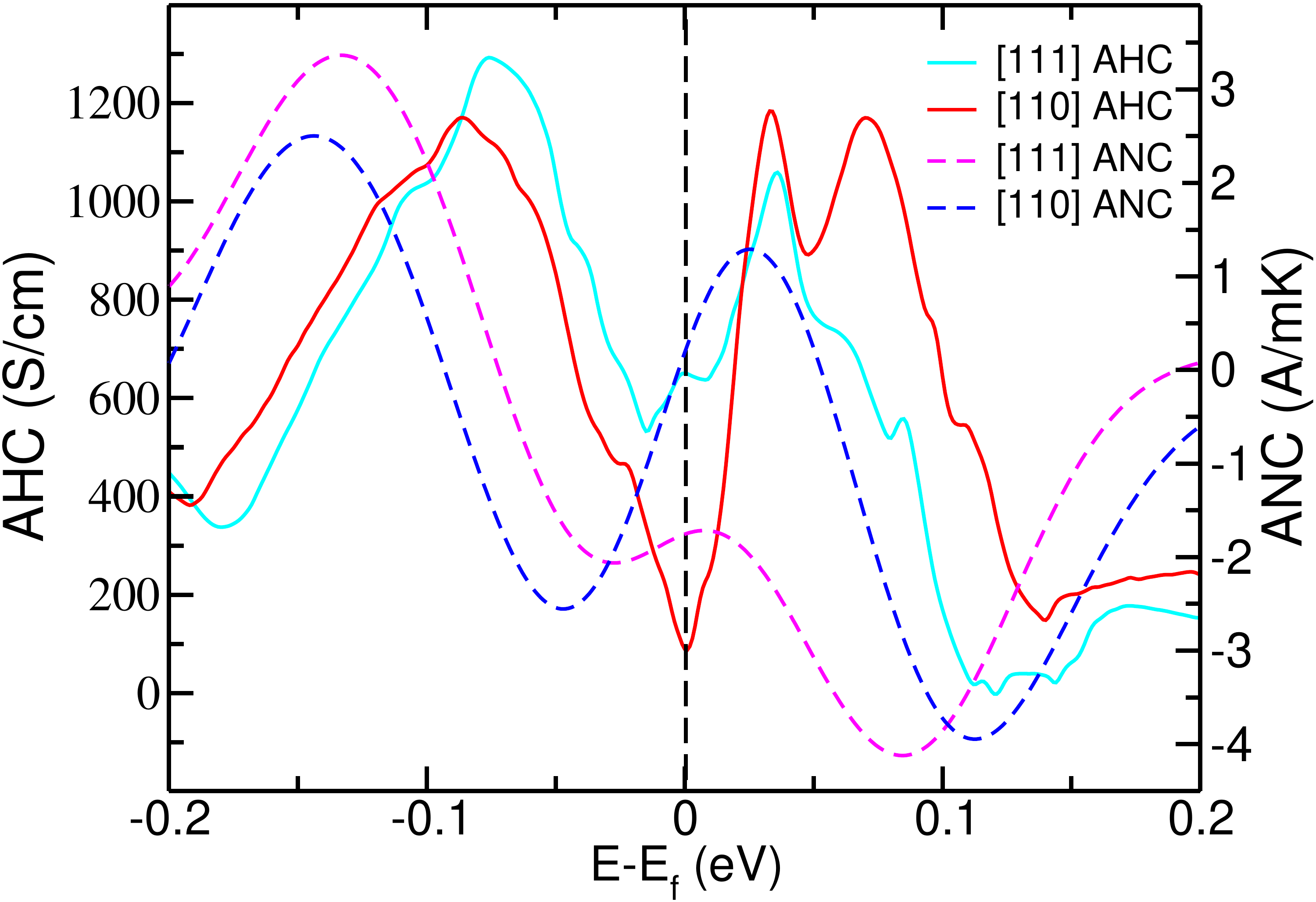}
		\caption{The AHC and ANC of Co$_3$LiN for [111] and [110] magnetization directions, exhibiting the significant change in AHC and ANC on switching the magnetization direction from [111] to [110].}
		\label{fig:CoLiN}
	\end{center}
\end{figure}
\vspace{1mm}
\subsection{Large Anomalous Hall Conductivity}
Figure~\ref{fig:AHC-ANC} displays the resulting AHC and ANC for the cubic APV compounds. The FM cubic APVs can attain an AHC as large as 1000 S/cm.
For instance, the magnitude of AHC for Co$_3$LiN and Co$_3$ZnN with magnetization aligned along the [111] direction is 1128 and 1069 S/cm, respectively. This is larger than that of bcc Fe (750 S/cm)~\cite{yao2004first,weischenberg2011ab} and nearly the same as that of the Weyl semimetal Co$_3$Sn$_2$S$_2$ (1130 S/cm)~\cite{liu2018giant} and Fe$_3$Sn$_2$ (1100 S/cm)~\cite{ye2018massive}. Moreover, the magnitude of AHC can be tuned significantly by changing the magnetization direction. 
For instance, the AHC of Co$_3$LiN changes from 1128 to 166 S/cm with the magnetization switched from the [111] to [110] directions (Figure~\ref{fig:CoLiN}). Furthermore, not only the magnitude of AHC can be tuned, but also its sign can be reversed by switching the magnetization directions. Taking an example of Co$_3$MgC, its AHC is 283 S/cm for the magnetization along the [001] direction, whereas it attains a negative value (-291 S/cm) for the [111] magnetization direction. Note that for high-symmetry magnetization directions such as [001], [110], and [111] directions considered in this work, both AHC and ANC are aligned along the direction of magnetization, where a positive (negative) sign indicates that they are parallel (antiparallel) to the magnetization direction. The same is observed for Co$_3$RhN, Co$_3$SnC, Co$_3$ZnC, Fe$_3$AlC, and Ni$_3$CuC (Table.~\ref{tab:table1}).
\begin{figure*}
	\begin{center}
		\includegraphics[width=1.0\textwidth]{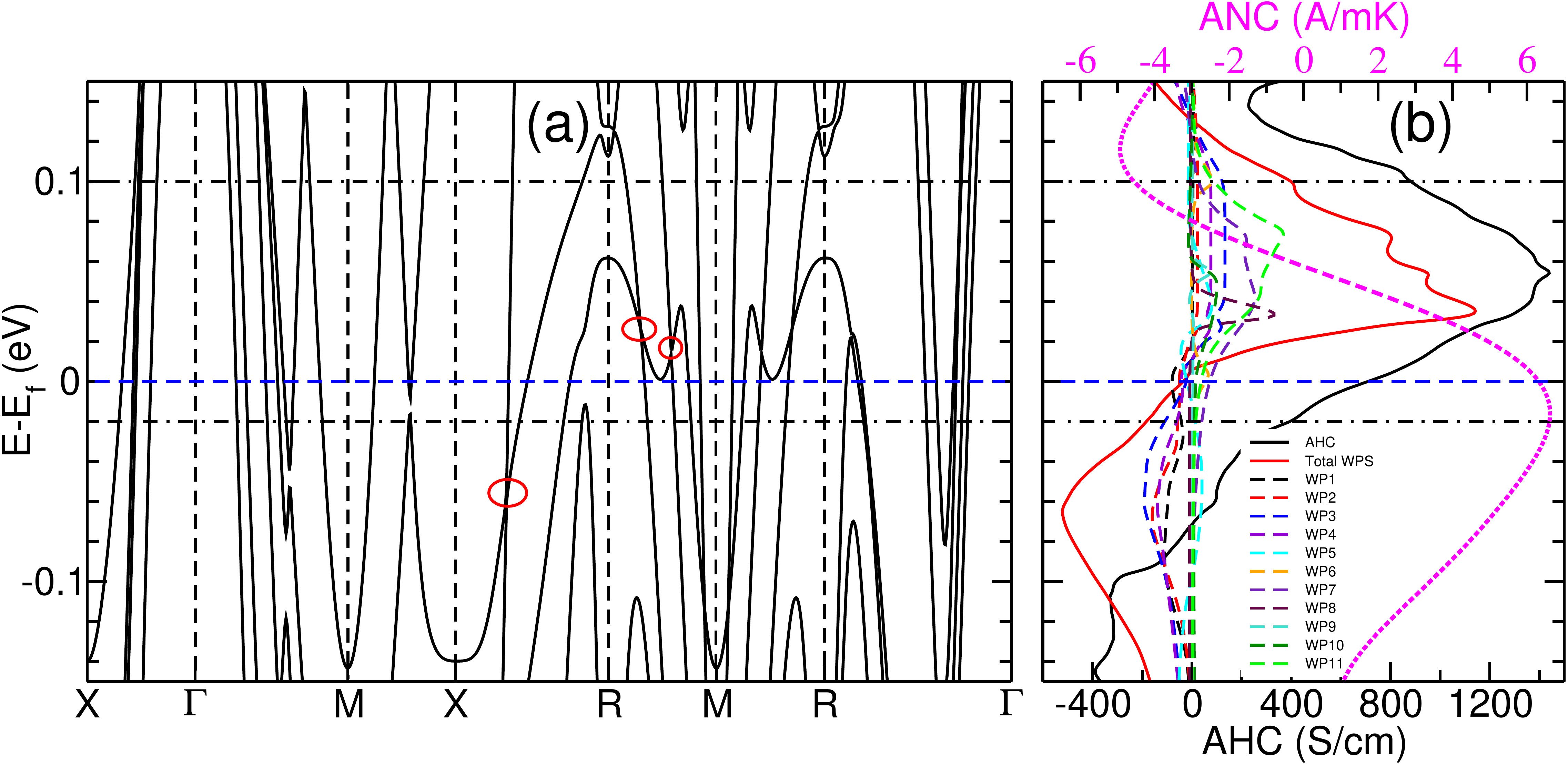}
		\caption{The calculated (a) band structure of Co$_3$PtN, where the red circles are pointing out the band crossing between R-M and X-R, and (b) Total WPs (red) is the sum of AHC from each symmetry-associated Weyl points (WP1-11), contributing mainly to the total AHC (black). For each symmetry-related WPs, the AHC is evaluated by considering a cube centered at spatial WP positions and acme of 0.08 (in units of 2$\pi$/a). The ANC is presented for 300 K (magenta).}
		\label{fig:AHC-CoPtN}
	\end{center}
\end{figure*}
\subsection{Comparison of AHC/ANC with noncollinear APVs}
It is noted that the magnetic cubic APV compounds exhibit many interesting magnetic ground states. In our recent study,~\cite{singh2020multifunctional} the total energies of eight magnetic configurations are evaluated and compared, where 14 cubic APVs adopt the noncollinear configurations of the $\Gamma_{4g}$/$\Gamma_{5g}$ types. The AHC/ANC vanishes for the $\Gamma_{5g}$ state, whereas a finite AHC/ANC is expected for the $\Gamma_{4g}$ state.~\cite{Samathrakis-Mn3GaN} As shown in Figure~\ref{fig:AHC-ANC}, the AHC of FM APVs is in general larger than that of those with noncollinear configurations. As an illustration, the largest AHC of 1128 S/cm is obtained for Co$_3$LiN in FM state, which is larger than the maximum AHC of 722 S/cm reported for noncollinear Cr$_3$IrN. In total, we found that six FM APVs exceed the maximum AHC (722 S/cm) observed for the noncollinear state.~\cite{singh2020multifunctional} Additionally, it is noted that the APV compounds are closely connected to those with the Cu$_3$Au-type crystal structures.~\cite{opahle2020effect} It is observed that giant AHC can be obtained there as well, such as in Mn$_3$Ir (218 S/cm)~\cite{chen2014anomalous} and Mn$_3$Pt (81 S/cm)~\cite{liu2018electrical}. 
\\
\subsection{Giant Anomalous Nernst Conductivity}
Interestingly, it is observed that such cubic APV compounds exhibit giant ANCs, which obey the same symmetry arguments as AHC. Following the Mott formula~\cite{xiao2006berry}, ANC is proportional to the energy derivative of AHC (the energy dependence of AHCs are summarized in Figure S3) and can be evaluated using Eq.~\ref{eq:ANC}. For FM APVs, Co$_3$PtN exhibits ANC as large as 6.27 AK$^{-1}$m$^{-1}$ at 300 K for the [001] magnetization direction (Table~\ref{tab:table1}). 
This is quite comparable to that of Fe$_3$Pt (6.2 AK$^{-1}$m$^{-1}$)~\cite{sakai2020iron} and larger than the experimentally observed ANC in the FM counterpart Co$_2$MnGa (4.0  AK$^{-1}$m$^{-1}$), Co$_2$FeGe (3.16 AK$^{-1}$m$^{-1}$), and Fe$_3$Ga (3.0  AK$^{-1}$m$^{-1}$).~\cite{sakai2018giant, guin2019anomalous, noky2018characterization} Furthermore, Ni$_3$LiC and Co$_3$ZnN also show a large ANC of 5.80 and -4.79 AK$^{-1}$m$^{-1}$ for the [111] magnetization direction, respectively. In total, seven APV systems attain an ANC larger than 3.0 AK$^{-1}$m$^{-1}$ at 300 K (Table~\ref{tab:table1}). 
\\
\\
Like AHC, ANC can also be tuned by switching the magnetization directions (Table~\ref{tab:table1}). For instance, the ANC of Co$_3$PtN alters from 6.27 AK$^{-1}$m$^{-1}$ to 3.79 AK$^{-1}$m$^{-1}$ as the magnetization direction changes from the [001] to [111] directions. Moreover, the sign of ANC could also change depending on the AHC slope's derivative d$\sigma$$_{xy}$/d$\epsilon$. For instance, the derivative of AHC for Co$_3$LiN changes from a positive (5318 S/cm.eV) value to a negative (-5190 S/cm.eV) one upon switching the magnetization direction from the [111] to [110] directions (Figure~\ref{fig:CoLiN}). Correspondingly, the resulting ANC changes from -3.02 to 0.55 AK$^{-1}$m$^{-1}$.
\\
\\
For Fe$_3$ZnN, the sign and magnitude of ANC change from -1.31 to 0.26 AK$^{-1}$m$^{-1}$ upon switching out-of-plane magnetization from [001] to [111]. Therefore, for an applied $\mathbf{M}$$_z$ $\parallel [001]/[111]$ and $\nabla$T$_x$ along the x-direction, an electric current will generate along -y/y direction (~\ref{fig:Thermopile}(a)). Similarly, for Co$_3$LiN, the sign of ANC changes from -1.73 to 0.55 AK$^{-1}$m$^{-1}$ for the ($\mathbf{M}$) [100] to [110]. When $\mathbf{M}$$_x$ $\parallel [100]/[110]$ is applied along the x-direction for the out-of-plane $\nabla$T$_z$, an electric current will generate along the -y/y-direction (~\ref{fig:Thermopile}(b)). Such an aspect is not only interesting in terms of the anisotropic nature of ANC, but also Nernst voltage change in sign could possibly result in the generation of in-plane $\nabla$T caused by rotation of magnetization direction. This type of behavior is demonstrated in a recent experiment study for the CoFeB/Ta sample, where in-plane rotation of magnetic field leads to the change in Nernst voltage sign; as a result thermal gradient generated in the sample.~\cite{lee2015thermoelectric} For the Fe$_4$N antiperovskite, Isogami et al. showed that the Nernst voltage could be tuned significantly by altering in-plane $\nabla$T or $\mathbf{M}$ directions from [100] to [110].~\cite{isogami2017dependence} Commonly, the ANE is studied in the experiment for the out-of-plane $\nabla$T and in-plane $\mathbf{M}$,~\cite{meier2013influence,kikkawa2013separation} and also for the in-plane $\nabla$T along with out-of-plane $\mathbf{M}$~\cite{tu2017anomalous,kikkawa2013separation}. However, the functional devices with out-of-plane $\nabla$T$_z$ are more beneficial in the context of practical application for heat harvesting, as illustrated for FePt thermopile with MgO(110) substrate.~\cite{sakuraba2013anomalous,sakuraba2016potential,mizuguchi2019energy}  
\\
\\
In comparison to the APVs with noncollinear magnetic ground states,~\cite{singh2020multifunctional} 
the ANC of FM APVs is more significant in most cases (Figure~\ref{fig:AHC-ANC}). 
For instance, the largest ANC obtained in noncollinear Cr$_3$PtN is 4.02 AK$^{-1}$m$^{-1}$, which is less than Co$_3$PtN. Overall, thirteen FM APVs attain an ANC larger than 2.0 AK$^{-1}$m$^{-1}$. In terms of applications of such compounds as transversal thermoelectric materials, one apparent advantage is that the resulting ANC can be switched by controlling the magnetization directions, as we discussed above, which is not so convenient for APVs with noncollinear magnetic ground states.
\subsection{Weyl point analysis}
\begin{figure}
	\begin{center}
		\includegraphics[width=0.49\textwidth]{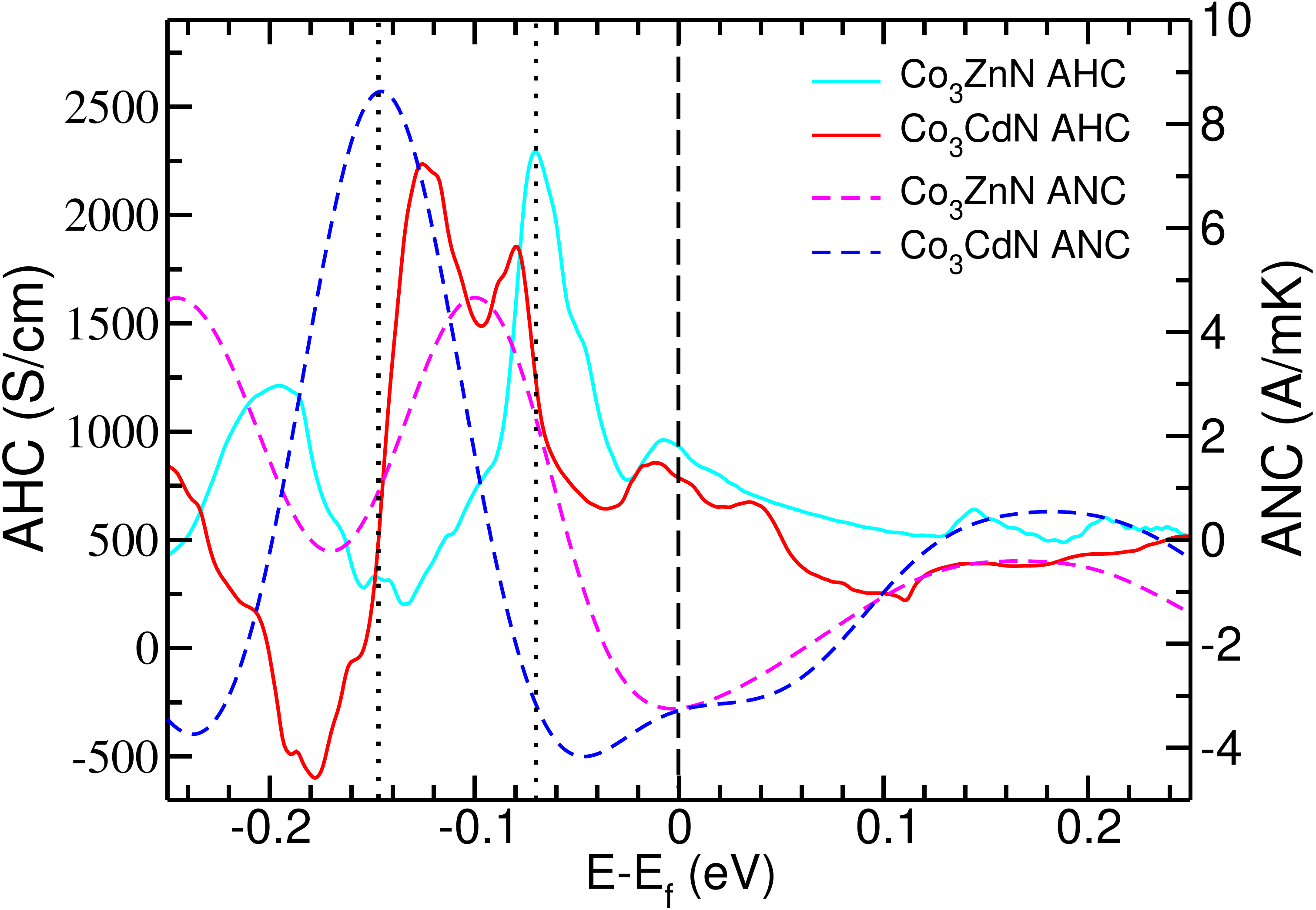}
		\caption{The AHC and ANC of Co$_3$ZnN and Co$_3$CdN for the [001] magnetization direction. A large AHC and ANC could be achieved by shifting the Fermi-energy 70 and 147 meV below Fermi-level, respectively.}
		\label{fig:AHC-ANC-CoZnN-CoCdN}
	\end{center}
\end{figure}

Turning now to the origin of giant ANC, it has been recently realized that the occurrence of Weyl nodes in the neighborhood of the Fermi-energy can contribute significantly to AHC and may lead to a singular-like behavior in AHC, thus a significant ANC.~\cite{wang2018large, shi2018prediction, xu2020high} Taking Co$_3$PtN with a giant ANC of 6.27 AK$^{-1}$m$^{-1}$ as an example, we carried out a detailed analysis of the electronic structure. As shown in Figure~\ref{fig:AHC-CoPtN}(b), there exists a peak of AHC located about 54 meV above the Fermi-energy. Correspondingly, there are two seemingly linear band crossings on the band structure, one above ($\sim$ 20 meV) the Fermi-energy along $R$-$M$ and the other below ($\sim$ 50 meV) the Fermi-energy along $X$-$R$ (Figure~\ref{fig:AHC-CoPtN}(a)). Further calculations reveal that there are 38 Weyl nodes within the [-0.02, 0.1]~eV energy window around the Fermi-energy, which can be classified into 11 sets, each related by the underlying symmetry. Notably, the sum of contributions to AHC from all such 38 Weyl nodes reproduces roughly the total AHC together with its
energy dependence around the Fermi-energy.(Figure~\ref{fig:AHC-CoPtN}(b)). 
Therefore, it is suspected that the giant ANC can be attributed to the occurrence of Weyl nodes. However, unlike Co$_3$Sn$_2$S$_2$, where there are three pairs of Weyl nodes located 60 meV above the Fermi-energy, and one pair of Weyl node acts as a monopole sink and source of Berry curvature,~\cite{liu2018giant} whereas for Co$_3$PtN, there are many of them, which is a general feature of FM materials.~\cite{gosalbez2015chiral} Similarly, the AHC peaks for Co$_3$LiN and Co$_3$ZnN (Fig.~\ref{fig:CoLiN} and~\ref{fig:AHC-ANC-CoZnN-CoCdN}) are also suspected to be resulting from the presence of Weyl nodes, but it is challenging to decompose the contribution because there are too many Weyl nodes. It is also noted that the occurrence of Weyl points does not necessarily always give rise to a singular-like AHC hence significant ANC.
\\
\subsection{Tuning AHC/ANC via chemical potential}
Such singular behavior of AHC does not occur only in the proximity to the Fermi-energy but also takes place at other energies. 
For instance, for the [001] magnetization direction, the AHC of Co$_3$ZnN can be tuned from 930 to 2249 S/cm by shifting the Fermi-energy by 70 meV below the Fermi level (Figure~\ref{fig:AHC-ANC-CoZnN-CoCdN}), with the ANC of 2.33 AK$^{-1}$m$^{-1}$.
Similarly, the ANC can be tuned, as for the Co$_3$CdN, an extremely large ANC 8.62 AK$^{-1}$m$^{-1}$ could be obtained at 147 meV below the Fermi-energy resulting due to presence of a sharp AHC peak which reaches to maximum AHC (2235 S/cm) at 126 meV below Fermi-energy (Figure~\ref{fig:AHC-ANC-CoZnN-CoCdN}). 
We suspect that such tuning can be accomplished by doping or mechanical strain, as demonstrated for the noncollinear Mn$_3$PdN APVs.~\cite{singh2020multifunctional} This will be saved for future studies.
\\ 
\section{CONCLUSIONS}
In summary, we carried out a systematic evaluation of MAE, AHC, and ANC for 35 cubic APVs. As expected for cubic systems, the MAE is few $\mu$eV/atom (0.2-10 $\mu$eV/atom) for most APV systems except a few cases; hence their magnetization direction can be tuned easily. A giant AHC as large as 1000 S/cm is obtained in cubic APVs such as Co$_3$LiN and Co$_3$ZnN. Similarly, a giant ANC is observed for Co$_3$PtN and Ni$_3$LiC. Taking Co$_3$PtN as an example, we demonstrated that a large AHC/ANC is originated due to the presence of Weyl nodes close to the Fermi-level, with dominant contributions to the total AHC/ANC. This indicates that AHC/ANC can be further tailored by shifting the Fermi-energy via proper doping. 
In comparison to noncollinear APVs, the FM APVs also exhibit larger AHC and ANC. Furthermore, we observed that the magnitude and sign of AHC/ANC could be tuned by switching the magnetization direction. Accordingly, we can adjust the thermal gradient ($\nabla$T) direction assisted by changing the magnetization directions. Therefore, it is imperative to perceive the importance of magnitude and sign change of ANC to realize ANC-based thermoelectric device applications. 
\section{ACKNOWLEDGMENTS}
The authors are grateful and acknowledge TU Darmstadt Lichtenberg's high-performance computer (HPC) support for the computational resources where the calculations were conducted for this project. The authors thank Dr. Benjamin Juhl from the TU Darmstadt HPC group for providing a special queue to perform the calculations. This project was supported by the Deutsche Forschungsgemeinschaft (DFG, German Research Foundation)-Project-ID 405553726-TRR 270.


\bibliography{my_file}
\bibliographystyle{ieeetr}

\end{document}